\documentclass{emulateapj}

\begin{document}


\title{Four New Stellar Debris Streams in the Galactic Halo}

\author{C. J. Grillmair}
\affil{Spitzer Science Center, 1200 E. California Blvd., Pasadena,  CA 91125}
\email{carl@ipac.caltech.edu}

\begin{abstract}

We report on the detection of four new stellar debris streams and a
new dwarf galaxy candidate in the Sloan Digital Sky Survey.  Three of
the streams, ranging between 3 and 15 kpc in distance and spanning
between 37\arcdeg~ and 84\arcdeg~ on the sky, are very narrow and are
most probably tidal streams originating in extant or disrupted
globular clusters.  The fourth stream is much broader, roughly 45 kpc
distant, at least 53\arcdeg~ in length, and is most likely the tidal
debris from a dwarf galaxy.  As the streams each span multiple
constellations, we extend tradition and designate them the Acheron,
Cocytos, Lethe, and Styx streams.  At the same distance and apparently
embedded in the Styx stream is a $\sim 1$ kpc-wide concentration of
stars with an apparently similar color-magnitude distribution which we
designate Bootes III. Given its very low surface density, its location
within the stream, and its apparently disturbed morphology, we argue
that Bootes III may be the progenitor of Styx and in possibly the
final throes of tidal dissolution. While the current data do not
permit strong constraints, preliminary orbit estimates for the streams
do not point to any likely progenitors among the known globular
clusters and dwarf galaxies.

\end{abstract}



\section{Introduction}

Despite the once common belief that the stellar debris streams
produced by tidal stripping of dwarf galaxies and globular clusters
would be quickly dispersed by molecular cloud scattering, orbital
precession, and phase mixing, recent observations of our Galaxy and
others have shown that such streams are both common and evidently
long-lived. The Sloan Digital Sky Survey (SDSS) has proven to be a
particularly remarkable resource for finding such streams, and for
studying Galactic structure at a level of detail which we
cannot hope to match in any other galaxy. In addition to the large
scale features attributed to past galaxy accretion events
\citep{yann03,maje2003,roch04,
  grill2006d, belokurov2006b, grill2006e,
  belokurov2007}, SDSS data has been used to detect the remarkably
strong tidal tails of Palomar 5 \citep{oden2001,rock2002, oden2003,
  grill2006b} and NGC 5466 \citep{belokurov2006a, grill2006a}, as
well as the presumed globular cluster stream GD-1 \citep{grill2006c}.
Though spectroscopic follow-up and detailed numerical calculations
have yet to be carried out for most of these streams, they will no
doubt become important for constraining the three dimensional shape of
the Galactic potential. Globular cluster streams will be particularly
important since they are dynamically very cold \citep{comb99} and
therefore useful for constraining not only the global Galactic
potential but also its lumpiness \citep{mura99}.

In this paper we continue our search of the SDSS database for more
extended structures in the Galactic halo. We describe our
analysis procedure in Section \ref{analysis}. We discuss four
new stellar streams and a possible dwarf galaxy progenitor in Section
\ref{discussion} and put initial constraints on their orbits in
Section \ref{orbit}. We make concluding remarks Section
\ref{conclusions}.

\section{Data Analysis} \label{analysis}

Data comprising $g,r,$ and $i$ photometry for $7 \times 10^7$ stars in
the region $108\arcdeg~ < \alpha < 270\arcdeg~$ and $-4\arcdeg~ < \delta
< 65\arcdeg~$ were extracted from the SDSS DR5 database using the SDSS
CasJobs query system. The data were analyzed using the matched filter
technique employed by \citet{grill2006a}, \citet{grill2006b}, \citet{
grill2006c}, \citet{grill2006d}, and \citet{grill2006e}, which itself is a
variation on the optimal filtering 
technique described by \citet{rock2002}. This technique is made
necessary by the fact that, over the magnitude range and over the
region of sky we are considering, the surface densities of foreground
stars are some three orders of magnitude greater than the surface
densities of known stellar debris streams.  Applied in the
color-magnitude (CM) domain, the matched filter is a means by which we can
optimally differentiate between streams and foreground populations.

Our filtering technique departs somewhat from that of
\citet{rock2002}, who were primarily interested in searching for
debris from a known and relatively well characterized progenitor. By
contrast, our present goal is to survey the sky for discrete but
hitherto unknown stellar populations. Since we are interested in
detecting streams throughout the Galactic halo, we also need to
account for the effects of survey completeness as we search larger and
larger volumes.  Consequently, rather than using the observed
color-magnitude distribution (CMD) for stars of interest (e.g.
\citet{rock2002}), we generate template distributions which are based
on the observed color-magnitude sequences of several globular clusters
situated within the SDSS survey area. Specifically, we measure normal
points lying along the $g-r$ and $g-i$ color-magnitude sequence of
each cluster and then interpolate to compute the expected color at any $g$
magnitude. Using mean photometric errors as a function of magnitude
(measured in relatively sparse regions of the survey area where source
crowding is not an issue) we broaden the globular cluster sequences by
convolving with appropriate Gaussians at each magnitude level. We also
apply a fixed broadening of $\sigma = 0.02$ mag at all magnitudes to
account for the intrinsic spread in the colors of giant branch stars.

Since we have no {\it a priori} knowledge concerning the luminosity
function of stars in streams, and since we need to decouple observed
luminosity functions from the survey completeness, we adopt a general
form for the luminosity function based on the very deep luminosity
function of $\omega$ Cen \citep{demarchi1999}, converted to the Sloan
system using the empirical transformations of \citet{jordi2006}. We
find that the exact form of the luminosity function is not
particularly important. Experiments with the somewhat steeper
luminosity functions one might expect for tidally stripped stars
\citep{koch2004} yield no perceptible improvements over the range of
absolute magnitudes considered here.

Comparing the observed luminosity function in the outskirts of M 13
with the much deeper $\Omega$ Cen luminosity function, we derive an
approximate completeness function which is unity at $g = 22$, 0.5 at
$g = 23.3$, and 0 at $g = 24.4$. While the actual completeness will
vary across the survey area, practical concerns require that we use a
single completeness function for the entire field. Since we impose a
cutoff at $g = 22.5$, small modifications to the form of
our completeness function have only very minor effects on the results.
Our expectation is that mismatches between our template
color-magnitude sequences and the largely unknown distributions of
stars in stellar streams will have a much larger effect on our
sensitivity to discrete populations.

Once the color-magnitude distribution of the stars of interest has
been constructed, an optimal filter requires that this distribution be
divided by the corresponding distribution of field stars (e.g.
\citet{rock2002}). We sample the field star distribution over various
portions of DR5, binning the stars in $g$ and $g-i$ (or $g-r$), and
then slightly smoothing over the bins with a Gaussian of kernel
$\sigma = 0.02$ mag. Figure 1 shows a template filter based on the CMD
of M 13 at its nominal distance of 7.7 kpc \citep{harris96}. To avoid
numerical issues in relatively unpopulated regions of the
color-magnitude diagram, we set the filter to zero for stars more than 
6$\sigma$ from the color-magnitude sequence. Examination of Figure 1
shows that the most highly weighted stars are those at the main
sequence turn-off. Stars fainter and redward of the turn-off are much
less favored, though their integrated contribution remains
significant. By virtue of both their relatively small numbers, and of
colors that are indistinguishable from the bulk of the foreground
population, the subgiant and lower giant branch stars are given
comparatively little weight.

\begin{figure}
\epsscale{1.0}
\plotone{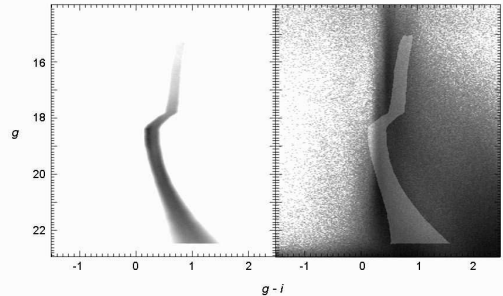}
\caption{Panel (a): An example of a matched filter based on the
SDSS color-magnitude distribution of stars in M 13, but with a luminosity
function based on $\Omega$ Cen. The stretch is logarithmic. The peak response of
the filter occurs on the blue side of the main sequence turn-off, where
there are relatively few foreground stars. The subgiant and lower giant branch are somewhat deweighted due to the very large number 
of intervening foreground stars in this color range. The width of the
filter is based primarily on the mean SDSS photometric errors as a function of
magnitude. Stars lying more than $6\sigma$ from the sequence are given a weight of zero. Panel (b): The color-magnitude distribution of stars lying within 10\arcdeg~ of the eastern half of the Cocytos stream, with the filter in panel (a) overlaid. \label{fig1}}
\end{figure}

We used all stars with $15 < g < 22.5$, and we dereddened the SDSS
photometry as a function of position on the sky using the DIRBE/IRAS
dust maps of \citet{schleg98}. In an initial survey, a single Hess
diagram for field stars was generated using roughly half the Sloan
survey area in regions where no streams are currently known to exist.
We applied the filters to the entire survey area, and the resulting
filtered star counts were summed by location on the sky to produce two
dimensional, filtered surface density maps. Once the streams were
detected, we optimized the filters for individual streams.  Since the
CM distribution of the field star population varies over the survey
area, we expect that a filter incorporating the distribution of only
nearby field stars will enhance the signal-to-noise ratio of the
streams. For each stream we sampled the field star population within
10\arcdeg~ of the stream, and with one exception, extending along only
the eastern and western halves of each stream. This has the effect of
increasing the measured signal-to-noise ratio by a few percent beyond
what one can achieve using a single, survey-wide field star
distribution. Panel (b) of Figure 1 shows an example of one such field
star distribution. Further improvements may be possible by more finely
subdividing the field star populations, or modeling the foreground
population as a function of sky position, but that is beyond the scope
of this paper.

In Figure 2 we show the filtered star count distributions after
shifting the optimal filters by -1.4, +0.6, +1.2 (M 13-based filter) and +3.2
mags (M 15-based filter).  The field area is shown in the Sloan Survey
coordinate system to improve visibility and reduce the distortions that
a projection in the equatorial or Galactic coordinate systems would
entail. The images have been binned to a pixel size of
$0.1\arcdeg~$ and smoothed using a Gaussian kernel with $\sigma =
0.2\arcdeg$.

\begin{figure*}
\epsscale{1.0}
\plotone{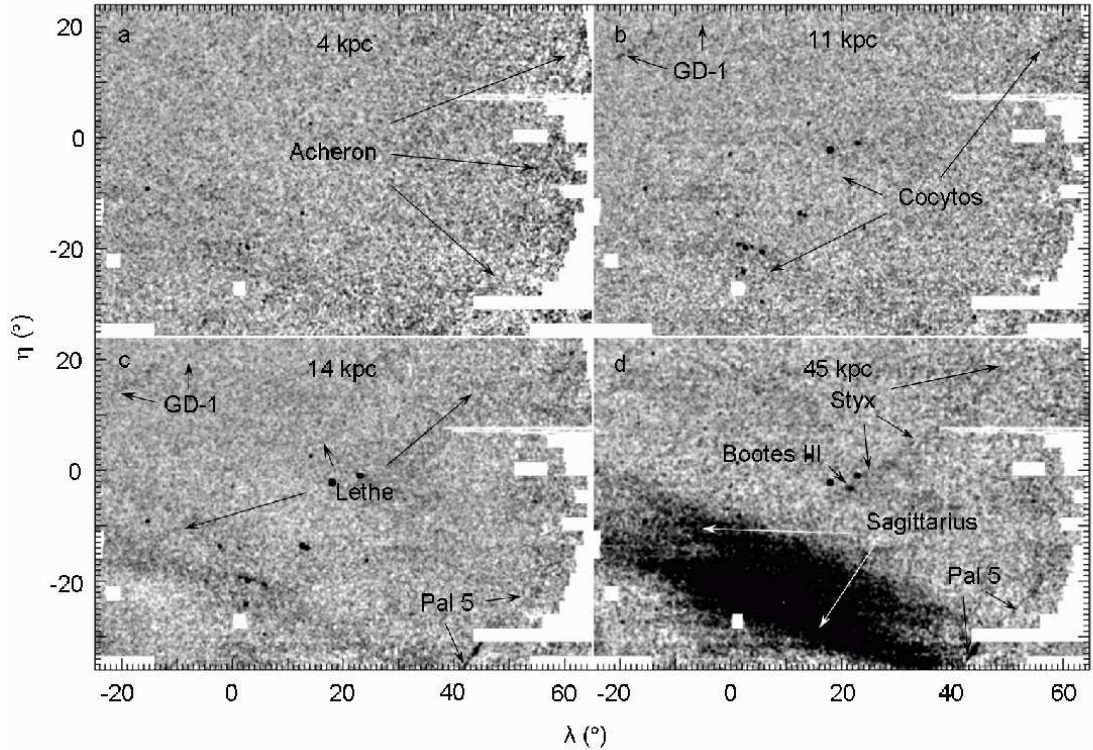}
\caption{Matched-filtered surface density maps of stars in the eastern 2/3rds of
the DR5 Sloan Digital Sky Survey field. The stretch is logarithmic,
and darker areas 
indicate higher surface densities. For panel (a) a 5th-order
polynomial fit has been subtracted from the surface densities for
presentation purposes. For the other three panels a 7th-order
polynomial surface fit has
been has been subtracted. All fields have been
smoothed with a Gaussian kernel of width 0.2\arcdeg~.  The white areas
designate areas of missing data. Panels (a) through (d) result from
shifting the M13 or M 15-matched filters by -1.4, +0.8, +1.1, and
+3.2 magnitudes, respectively, and the corresponding distances
(assuming $d_{M13} = 7.7$ kpc and $d_{M15} = 10.3$ kpc) are
indicated. \label{fig2}}
\end{figure*}

The filtered surface density maps are the sum of maps generated using
$g - r$ and $g - i$ filters as these colors best measure the turn-off
and main sequence stars of interest.  To improve the visibility of the
streams in Figure 2, each image has been background subtracted by
first masking out globular clusters and dwarf galaxies and then
fitting a 5th or 7th order polynomial surface.  These surface fits are
shown in Figure 3.  For the nearest of the streams, a 5th order
polynomial fit is found to be sufficient to remove the rise in the
number of disk stars at low Galactic latitudes. For the remaining
streams a 7th order polynomial was used to subdue the increasing
contribution from the Sagittarius stream. As is
evident in Figure 3, there are no high-frequency features in the
surface fits that could could be held to account for the streams
visible in Figure 2. We emphasize that these background subtractions
are purely for the purposes of presentation and we make no further use
of them in our subsequent analysis.

\begin{figure}
\epsscale{1.0}
\plotone{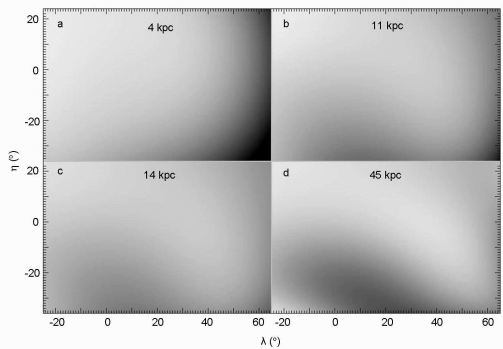}
\caption{Polynomial surface fits used to remove the background in
Figure 2. Panel (a) is the result of using a 5-th order polynomial
fit, while all 
other panels employ a 7th-order polynomial to reduce the effect of the
Sagittarius stream. None of the streams in Figure 2 can be associated
with the much lower frequency undulations visible here. \label{fig3}}
\end{figure}

We compared Figure 2 with the reddening map of \citet{schleg98}
to ensure that apparent stellar over-densities are not due to
localized changes in extinction. The reddening map covering the field
of interest is shown in Figure 4. There is no apparent correlation between
the new streams and the applied reddening corrections. The maximum
values of $E(B-V)$ in the regions subtended by the new streams are
$\sim 0.13$, with typical values of $< 0.05$. Rerunning the matched
filter analysis without reddening corrections has no significant 
effect on the location or the apparent strengths of the new features.

\begin{figure}
\epsscale{1.0}
\plotone{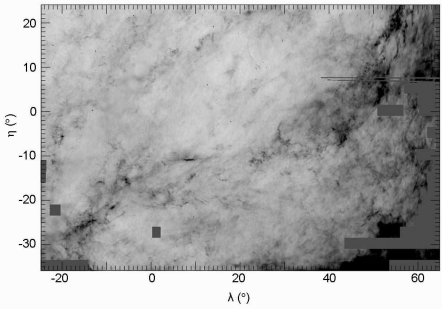}
\caption{$E(B-V)$ over the region of sky shown in Figure 2, as determined from the
the DIRBE/IRAS dust maps of \citet{schleg98}. The stretch is linear,
with darker regions corresponding to higher reddening. The reddening varies from $< 0.05$ mag to 0.13 mag at the
locations of the four streams. \label{fig4}}
\end{figure}

We have also compared Figure 2 with similar maps made using the
SDSS DR5 galaxy database to ascertain the extent to which confusion
between stars and galaxies at faint magnitudes could contribute
to the features we see. In none of the four cases presented here is
there any indication of similar extended features in the distribution of
galaxies.

\section{Discussion} \label{discussion}

Visible in Figure 2 are several well known tidal features, as well as
four new stellar debris streams. The new streams are much
less pronounced than the Sagittarius, Pal 5, or GD-1 streams, and have
average surface densities of between 5 and 50 stars deg$^{-2}$ The
streams were initially detected and are most easily distinguished by
viewing a rapid sequence of filtered images in which the filter is
successively shifted to fainter and fainter magnitudes. The streams
become apparent to the eye as linear features which often
move from one side of the survey area to the other as one moves
outward in distance.

All of the new streams span multiple constellations and none can be
securely identified with a known progenitor at this time. For
convenience, we extend traditional nomenclature and name the
new streams after four mythical rivers in Hades: the Acheron (river of
sorrow), Cocytos (lamentation), Lethe (forgetfulness), and Styx (hate).

\subsection{Acheron \label{acheron}}

Visible in panel (a) of Figure 2 is a fairly narrow stream of stars
extending some $37\arcdeg~$ across the southeastern corner of the
DR5 survey area. The stream extends from the southern edge of Serpens
Caput [($\lambda, \eta$) = (45\arcdeg, -36\arcdeg), (R.A., dec.) =
(230\arcdeg, -2\arcdeg)] to the center of Hercules [($\lambda, \eta$) =
(63\arcdeg, 22\arcdeg), (R.A., dec.) = (259\arcdeg, 21\arcdeg)], and is
truncated at both the southern and eastern ends by the limits of the
available data.

The stream is much less pronounced than (for example) GD-1
\citep{grill2006c} and in places the signal-to-noise ratio is almost
vanishingly small. To better quantify the significance of the
detection, we apply the following test: (i) We trace along the length
of the putative stream, connecting the high points in the surface
density distribution with segments that match the general curvature of
the stream (though in this case there is almost none).  (ii) We create
a mask image of this trace, setting pixels along the trace and
laterally out to 0.25\arcdeg~ in each direction to unity.  All other
mask pixels are set to zero. This width is chosen to roughly match the
apparent width of the stream.  (iii) We break the mask into seven
stream segments, each approximately 5\arcdeg~ in length.  (iv) For each
stream segment, we successively shift the mask in the $\lambda$
direction across a representative portion of the sky, 0.1\arcdeg~ at a
time, and multiply the mask by the filtered image in Figure 2. (v) For
each segment, we fit a one dimensional, 3rd order polynomial to the
mean filtered star counts as a function of lateral distance from the
stream and subtract it. We exclude from this fit the region within
0.5\arcdeg~ of the stream.  (vi) For each lateral offset, we compute
the median of the background-subtracted responses over all segments.
This last calculation serves as a continuity constraint and prevents
strong biasing due to (for example) a single populous star cluster in
in any one segment.

More succinctly, if we define $f(i,j)$ as the filtered star counts in
a pixel with indices $i$ and $j$, and $m(i,j)$ as the stream-tracing
mask, then the signal $s$ for stream segment $k$, shifted laterally by
offset $d$, is:

\begin{equation}
s(d,k) = \frac{\sum_{i,j}[f(i,j) \times m(i+d,j)]}{ n(d,k)},
\end{equation}

where the summation is over all valid pixels (i.e. excluding
portions of the Survey footprint with missing data). The total stream
signal at offset $d$ is then:

\begin{equation}
T(d) = {\rm median} [s(d,k)]_{k=1,...l},
\end{equation}

where $l$ is the number of stream segments.

We plot the run of $T$ versus lateral offset for Acheron in Figure 5.
The signal due to the stream is clearly visible as a broad peak
approximately centered on a lateral offset of 0\arcdeg. We note that
perfect centering is not expected; aside from measurement
uncertainties in regions with very little signal, the lateral profiles
of real tidal streams need not be symmetric, depending on such factors
as our line of sight to the stream and what portion of the stream's
orbit is being observed.  If we regard $T$ beyond 1\arcdeg~ from the
stream as being due to random clumping of stars and therefore
a reasonable measure of the noise, we find $\sigma_{T} \approx 5$,
where the standard deviation is measured after binning over the
0.5\arcdeg~ width of the mask.  Integrating over the region $-1\arcdeg
< d < 1\arcdeg$ we find that we have detected the stream at the $\sim
10 \sigma$ level. For comparison, a similar test applied to GD-1
yields a signal-to-noise ratio of $\sim 13$.

\begin{figure}
\epsscale{1.0}
\plotone{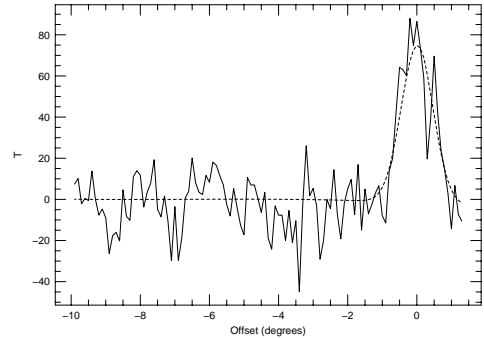}
\caption{Stream signal $T$ (described in the text) as a function of $\lambda$ offset from the Acheron stream. The dashed line shows the results of a similar test using an artificial stream with a Gaussian cross section and 
FWHM = 0.9\arcdeg. \label{fig5}}
\end{figure}

Also shown in Figure 5 is an identical test applied to an artificial
tidal stream at the same location as Acheron and with a Gaussian cross
section. The Gaussian which yields the best match to the observed
stream profile has a full width at half maximum (FWHM) of 0.9\arcdeg.
While the physical cross section of the stream need not be Gaussian,
the convolved artificial stream profile matches the actual stream
profile reasonably well, and we use the best fitting FWHM as a
convenient measure of the stream's breadth. At a distance of 3.6 kpc
(see below), this corresponds to a physical width of 60 pc. This is
similar to the widths measured for the tidal tails of the globular
clusters Pal 5 and NGC 5466
\citep{grill2006b,belokurov2006a,grill2006a} and the presumed cluster
remnant GD-1 \citep{grill2006c}. On the other hand, the width is much
narrower than the tidal arms of the Sagittarius dwarf
\citep{maje2003,mart2004} or the presumed dwarf galaxy streams
discussed by \citet{grill2006d}, \citet{belokurov2006b},
\citet{belokurov2007}, and \citet{grill2006e}. This is consistent with
the hypothesis that the stars making up the stream have very low random
velocities, and that they were weakly stripped from a relatively small
potential. Combining this with an orbit which passes low over the
Galactic bulge (see below) suggests that the parent body is (or was) a
globular cluster.

Following \citet{grill2006c} we shift the main sequence of the M
13-based filter brightward and faintward to estimate the stream's
distance. To avoid potential problems related to a difference in age
between M 13 and the stream stars, we use only the portion of the
filter with $19.5 < g < 22.5$, where the bright cutoff is 0.8 mags
below M 13's main sequence turn-off. This significantly reduces the
contrast between the stream and the background (since turn-off stars
contribute a substantial portion of the signal) but still provides
sufficient signal to enable a reasonably precise measurement of peak
contrast. We find that the strength of the southern end of the stream
peaks at a magnitude shift of -1.53 mags, the central portion has the
highest contrast at -1.58 mags, and the northernmost portion of the
stream peaks at -1.73 mags. Adopting a distance to M 13 of 7.7 kpc
\citep{harris96}, this puts the southern end of the stream at a
heliocentric distance of 3.8 kpc, while the northern end is at 3.5
kpc. While the match between the color-magnitude distributions of
stars in M 13 and in the stream is uncertain, the {\it relative}
line-of-sight distances along the stream should be fairly robust; 
we estimate our random measurement uncertainties to be $\approx 10\%$.

{\it Absolute} distance estimates using this method depend not only on
the uncertainty in the distance to M 13 but also on the CMD of
foreground stars and the degree to which the metallicity (and hence
color) of M 13's main sequence matches that of the stream. We have
only a very coarse estimate of the latter, namely the maximum contrast
obtained for the stream when the star counts are processed with
matched filters made from different globular clusters. If as a rough
estimate of this uncertainty we take half the $g$ magnitude offset
(0.46 mag) at a fixed color between the (dereddened) main sequence
loci of M13 and M 15 (with [Fe/H] of -1.54 and -2.25, respectively),
and combine this with a 5\% uncertainty in the distance to M 13
\citep{grundahl1998}, we arrive at a probable lower bound on the
systematic distance uncertainty of 11\%.

Integrating the locally background subtracted, filtered star counts
over a width of $\approx 1\arcdeg~$ we find the total number of stars
in the discernible stream to be $1300 \pm 200$. For stars with $g <
22.5$ the average surface density is $50 \pm 5$ stars deg$^{-2}$, with
occasional peaks of over 100 stars deg$^{-2}$. While these surface
densities are similar to those found by \citet{grill2006c} for GD-1,
Acheron appears considerably less pronounced. This is simply a
consequence of the much larger number of contaminating foreground
stars near the Galactic plane.

\subsection{Cocytos \label{cocytos}}

Visible in panel (b) of Figure 2 is a faint, narrow stream extending
from Virgo [($\lambda, \eta$) = (1\arcdeg, -35\arcdeg), (R.A., dec.) =
(186\arcdeg, -3\arcdeg)] in the south to Hercules [($\lambda, \eta$) =
(65\arcdeg, 20\arcdeg), (R.A., dec.) = (259\arcdeg, 20\arcdeg)] in the east.
The 80\arcdeg~~ length of the stream is again limited by the extent of
the SDSS survey area.

In Figure 6 we show the run of $T$ with lateral distance from the stream.  In
this case we have shifted our stream mask at a 45\arcdeg~ angle across
Figure 2 and divided the stream mask into 12, $\sim 5\arcdeg$-long
segments. We find that $\sigma_{T} \approx 5.0$ at $|d| > 1\arcdeg~$ and,
integrating from $-1\arcdeg < d < 1\arcdeg$, determine that we have
detected the stream at the $\sim 8\sigma$ level. Generating an artificial
stream and applying the same test, we find that we obtain the best
match to the observed profile using a Gaussian with FWHM =
0.7\arcdeg~. At a distance of 11 kpc (see below) this corresponds to a physical
width of 140 pc. This is again similar to known globular cluster
streams and argues that the progenitor of Cocytos is (or was) a
globular cluster.

\begin{figure}
\epsscale{1.0}
\plotone{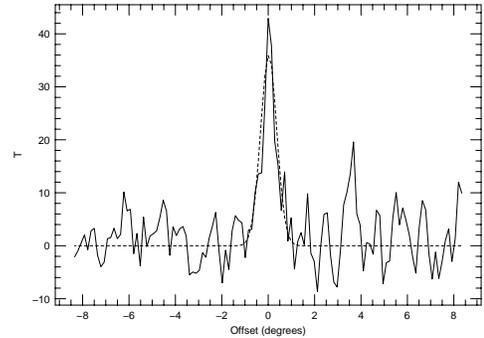}
\caption{As in Figure 5, but for the Cocytos stream. The artificial
stream that best matches the observed $T$ profile has a Gaussian
profile with FWHM = 0.7\arcdeg. \label{fig6}}
\end{figure}

The strength of both the southern and eastern ends of the stream peak
at an M 13 main sequence magnitude shift of 0.9 mags.  The estimated
distance to the stream is therefore $11 \pm 2$ kpc.  Integrating the
background subtracted, weighted star counts over a width of $\approx
1\arcdeg~$ we find the total number of stars in the discernible stream
to be $500 \pm 100$. For stars with $g < 22.5$ the average surface
density is between 5 and 8 stars deg$^{-2}$.

\subsection{Lethe \label{lethe}}

Panel (c) of Figure 2 shows a faint stream extending from
Hercules [($\lambda, \eta$) = (64\arcdeg, 19\arcdeg), (R.A., dec.) =
(258\arcdeg, 20\arcdeg)] to Leo [($\lambda,\eta$) = (-13\arcdeg,
-14\arcdeg), (R.A., dec.) = (171\arcdeg, 18\arcdeg)]. The stream extends
from the eastern limit of the survey and appears to fade substantially
in Leo, where it crosses the Sagittarius stream. There may be a
continuation of the stream south of the Sagittarius stream but we
have been unable to identify it with any confidence.

The apparent strength of Lethe relative to the distribution of
foreground stars peaks at an M 13-relative offset of 1.3 mags at the
eastern end, and about 1.0 mag at the western end. This puts the stream at
a distance of between 12.2 and 13.4 kpc. Figure 7 shows the run of
$T(d)$ for the stream, where again we have translated the stream mask at a
45\arcdeg~ angle across the filtered imaged in Figure 2 and subdivided
the mask into 12 stream segments. In this case we find $\sigma_{T} =
1.8$ at $|d| > 1\arcdeg~$, and integrating within this region we find
that Lethe is detected at
the $7\sigma$ level.  The best matching Gaussian stream shown in Figure 7
has a FWHM of 0.4\arcdeg, which at 13 kpc corresponds to a physical
width of 95 pc.  Once again we conclude that Lethe is the debris
stream of a globular cluster.

\begin{figure}
\epsscale{1.0}
\plotone{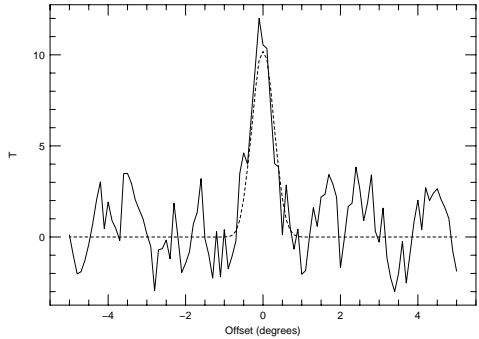}
\caption{As in Figure 5, but for the Lethe stream. The artificial
stream that best matches the observed $T$ profile has a Gaussian
profile with FWHM = 0.4\arcdeg. \label{fig7}}
\end{figure}

Integrating the background subtracted, weighted star counts over a
width of $\approx 1\arcdeg~$ we find the total number of stars in the
discernible stream to be $1100 \pm 300$. For stars with $g < 22.5$
the average surface density is 12 stars deg$^{-2}$, with the more
pronounced regions of the stream having $\sim 30$ stars deg$^{-2}$.

\subsection{Styx}  \label{styx}

Visible in panel (d) of Figure 2 is a broad stream extending west from
Hercules [($\lambda, \eta$) = (63\arcdeg, 21\arcdeg), (R.A., dec.) =
(259\arcdeg, +21\arcdeg)] to where it is overwhelmed by the much more
populous Sagittarius stream in Coma Berenices [($\lambda, \eta$)
= (8\arcdeg, -12\arcdeg), (R.A., dec.) = (194\arcdeg, +20\arcdeg)].
The contrast for Styx is improved if, instead of using a matched
filter based on M 13, we use the SDSS CMD of M 15. This
suggests that the CMD of stars in Styx is bluer than that of M
13.  At the distance of Styx the stellar main sequence is almost
entirely beyond the SDSS limiting magnitude. Consequently the main
sequence fitting technique for estimating distance is not
usable and we are forced to rely on turn-off and subgiant stars. Using
the M 15 filter, the apparent contrast of the stream peaks at
magnitude offsets of +2.8 mag at the western end, +3.2 mag in the
central portion, and +3.4 at the eastern end. Adopting a distance to M
15 of 10.3 kpc \citep{harris96}, this translates to estimated
distances of 38, 45, and 50 kpc, respectively. Owing to both the lack
of a direct main sequence comparison and to the rather distended
nature of the stream (which makes foreground estimation problematic),
we regard these distances as very approximate until such time as 
deeper photometry becomes available.

Figure 8 shows the run of $T$ with lateral offset. For Styx we have
used a mask width of 1\arcdeg, divided the mask into 12, $\sim
4\arcdeg$-long segments, and shifted the mask solely in the $\eta$
direction. In this case we find $\sigma_{T} = 0.18$ at $|d| >
3\arcdeg~$, and integrating from -3\arcdeg~ to +3\arcdeg~ we find that
Styx is detected at the $\sim 13\sigma$ level. The stream profile is
noticeably asymmetric, with a fairly sharp northern edge but an excess
of stars extending some 4\arcdeg~ to the south. This is qualitatively
similar to the morphologies of tidal tails in N-body simulations
(e.g. \citet{choi2007}). The artificial stream which best matches the
primary component of Styx has a FWHM of 3.3\arcdeg. At a distance of
45 kpc this corresponds to 2.6 kpc. This is much broader than the
presumed globular cluster streams above, but is similar to the the
dwarf galaxy streams discussed by \citet{grill2006d},
\citet{belokurov2007}, and \citet{grill2006e}, though narrower by half
compared to the Sagittarius stream
\citep{maje2003,mart2004,belokurov2006b}. We conclude that this stream
is most likely debris from a dwarf galaxy.

\begin{figure}
\epsscale{1.0}
\plotone{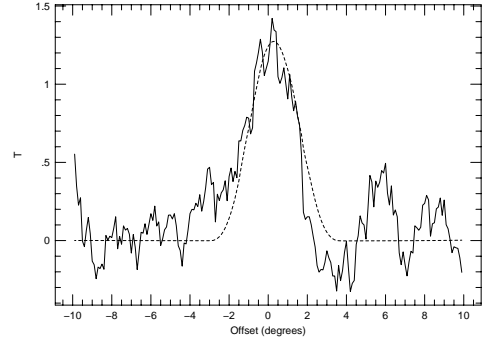}
\caption{As in Figure 5, but for the Styx stream. The artificial
stream that best matches the observed $T$ profile over the region $
-3\arcdeg < d < 3\arcdeg$ has a Gaussian
profile with FWHM = 3.3\arcdeg. The stream is
clearly asymmetric, with an extended spray of stars to the south, and
a fairly sharp cut off on the northern side.\label{fig8}}
\end{figure}

\subsection{Bootes III: A New Dwarf Galaxy?} \label{bootes3}

With a contrast maximum at very nearly the same distance ($\sim 46$
kpc) as the Styx stream is a relatively compact feature at [($\lambda,
\eta$) = (21.6\arcdeg, -3.5\arcdeg), (R.A., dec) =(209.3\arcdeg,
26.8\arcdeg)].  The object has a filtered star count surface density
many times higher than any visible portion of Styx and is clearly
distinct within the stream.  Panel (a) of Figure 9 shows the filtered
star count distribution in the immediate vicinity of this object.
Galaxy cluster ACO 1824 lies within 3\arcmin~ of the location of this
object\citep{abell1989}; could the apparent over-density be due to
SDSS misclassification of stars at faint magnitudes? In panel (b) of
Figure 9 we show the distribution of objects classified as galaxies in
DR5, where we have used a filter identical to that used in panel (a).
There is little correspondence between the two distributions, and
certainly no concentration of objects at the same position. We
conclude that this over-density is due to an equidistant collection of
stars orbiting in the outskirts of our own Galaxy and we henceforth
designate it Bootes III.

\begin{figure}
\epsscale{1.0}
\plotone{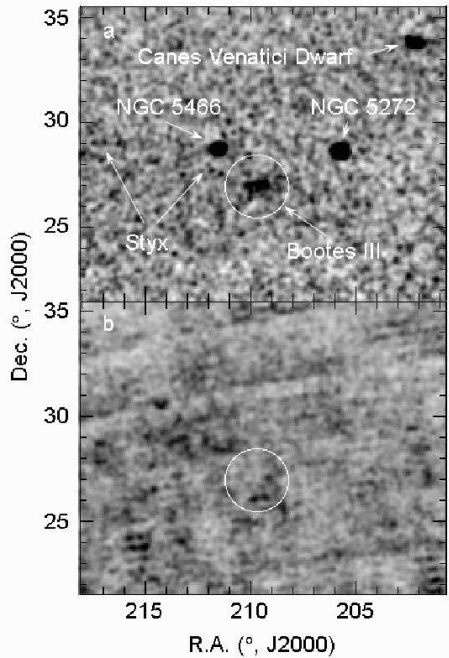}
\caption{(a) Matched-filtered surface density map in the immediate vicinity of Bootes
III, smoothed with a Gaussian kernel of width 0.1\arcdeg,  at an M
15-relative magnitude shift of 3.2 mags. Darker areas indicate higher
surface densities and the stretch is linear.  Though at very
different distances, NGC 5466, NGC 5272, and the Canes 
Venatici dwarf \citep{zucker2006} are also visible by virtue of their
high surface densities and the overlap of parts of their CMDs with 
the M 15 filter. (b) The distribution of objects classified as
galaxies in DR5, where we have used a filter identical to that used in
panel (a).  \label{fig9}}
\end{figure}

Figure 10 shows a contour plot of the filtered star counts in and
around Bootes III; the object appears somewhat double-lobed, possibly
disturbed, and extends $\approx 1.5\arcdeg~$ from east to west. At 46
kpc this corresponds to a spatial extent of about $\sim 1$ kpc. If the
stellar populations in the eastern and western lobes of Bootes III are
identical, then the difference in the filter shifts required to
maximize the apparent contrast indicates that the eastern lobe of the
galaxy is some 3 kpc closer to us than the western lobe.  If the two
lobes are indeed part of the same structure then Bootes III must be
highly elongated along the line of sight.  Figure 11 shows a surface
density profile of Bootes III, where we have counted all stars with $g
< 22.7$ and $-1 < g-i < 1$. The galaxy is evidently quite extended,
with a power-law surface density profile that goes as $\sigma \propto
r^{-1.0 \pm 0.2}$.  Integrating the background-subtracted counts out
to 1\arcdeg~ we find a total of 302 stars with $g \le 22.7$ which we
can reasonably attribute to Bootes III.

\begin{figure}
\epsscale{1.0}
\plotone{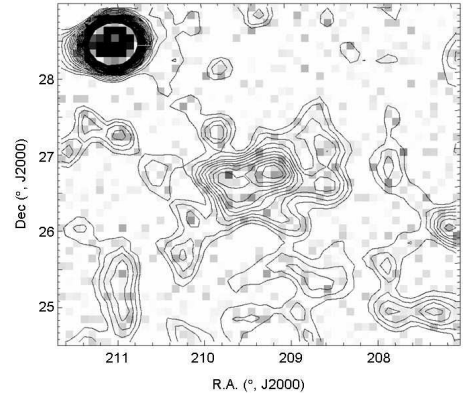}
\caption{Contour plot of a region centered on Bootes III. The underlying image has been smoothed with a Gaussian kernel of 6\arcmin. Contours are
spaced linearly, and the strong source to the northeast
of Bootes III is NGC 5466.} 
\end{figure}

\begin{figure}
\epsscale{1.0}
\plotone{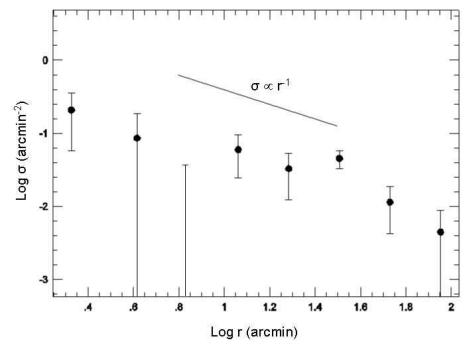}
\caption{The (unfiltered) surface density profile of all stars with $g < 22.7$ and $-1 < g -i < 1$. The center of Bootes III is taken to be at (R.A., dec) = (209.281\arcdeg, 26.775\arcdeg). The surface densities have been background subtracted using the measured counts in an annulus extending from 1.5\arcdeg~ to 2\arcdeg.}
\end{figure}

As is the case for Styx, using the SDSS color-magnitude distribution
of stars in M 15 as the basis for the matched filter yields a slightly
higher contrast between the galaxy and the foreground population. The
inference is that the age and/or metallicity of Bootes III and the
stream are more similar to that of M 15 than M 13. At 46 kpc the
galaxy is revealed almost entirely by subgiant and turn-off stars;
removing the red giant branch from the filter has little effect on the
apparent contrast. Figure 12 shows the color-magnitude distribution of
all stars within 1\arcdeg~ of the center of Bootes III. Here we have
subtracted the distribution of stars in several regions around Bootes
III over an area of $\sim 60$ square degrees. Comparing with the CM
loci of M 13 and M 15 (shifted vertically to a distance of 46 kpc),
there is a clear overdensity of stars along the expected positions of
the main sequence turn-off, the subgiant branch, and the lower giant
branch.  Consistent with the filter responses above, the CM locus of M
13 is evidently $\sim 0.1$ mag too red to match the apparent
distribution. The M 15 locus clearly provides a better match to
the data.

\begin{figure}
\epsscale{1.0}
\plotone{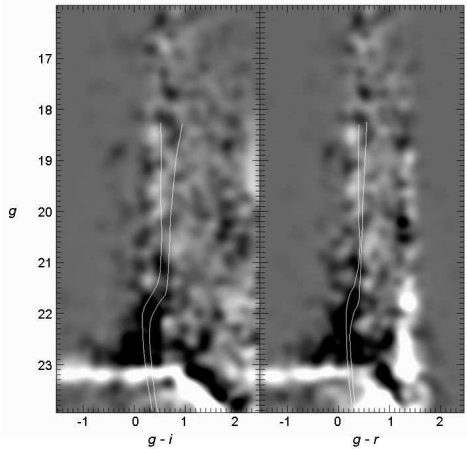}
\caption{The background-subtracted, color-magnitude distribution of
all stars lying within 1\arcdeg~ of the center of Bootes III. Darker
areas indicate enhancements over the background distribution. Stars
were binned in color and magnitude and smoothed with a Gaussian kernel of
$\sigma = 0.1$. The distribution was background-subtracted using
similarly binned and smoothed color-magnitude distributions in 
regions spaced around Bootes III and the two nearby globular clusters
and covering a total of $\sim 60$ square degrees. Also shown are the
$g, g-i$ and $g, g-r$ color-magnitude loci for M 13 (to the red) and M
15 (to the blue), as derived from SDSS photometry.}
\end{figure}

In Figure 13 we show the CMD of all stars within 0.8\arcdeg~ of the
center of Bootes III. While the turn-off, subgiant, and red giant
branches of Bootes III are lost among the unsubtracted foreground
stars in this figure, there is a clear concentration of stars at the
expected location of the blue horizontal branch (BHB). Fitting the BHB
sequence tabulated for the SDSS system by \citet{sirko2004}, we find
$(M-m)_0 = 18.35 \pm 0.01$. Corresponding to a distance of 46.7 kpc,
this is in excellent agreement with our maximum contrast distance
estimate above.  In Figure 14 we show the distribution of stars
selected to have colors and magnitudes consistent with Bootes III's
BHB. There is an apparent enhancement of such stars across the face of
Bootes III, though just as for the turn-off stars sampled using the
matched filter, the distribution of BHB stars is not very centrally
concentrated.  The BHB star distribution is considerably more extended
in the east-west direction (much like the filtered star counts in
Figures 9 and 10), subtending nearly 2\arcdeg on the sky.

\begin{figure}
\epsscale{1.0}
\plotone{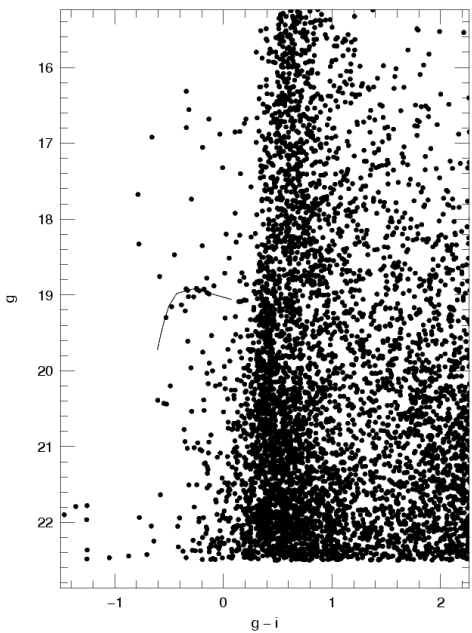}
\caption{The color-magnitude distribution of stars within 0.8\arcdeg~
  of the center of Bootes III. A blue horizontal branch is clearly
  visible at $(g, g - i) \approx 19, -0.25$). The solid line shows the
  predicted SDSS blue horizontal branch sequence of \citet{sirko2004},
  shifted vertically by 18.35 mags.}
\end{figure}

\begin{figure}
\epsscale{1.0}
\plotone{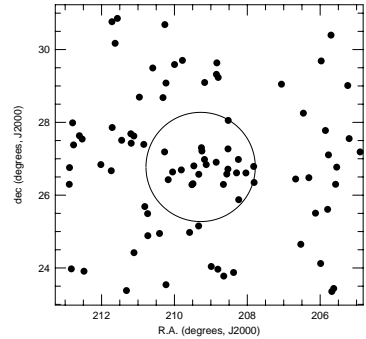}
\caption{The sky distribution of candidate Bootes III blue horizontal branch   stars. Based on the CMD in Figure 13, the stars are selected to have $18.7 < g < 19.3$ and $-0.55 < g - i < -0.07$.} 
\end{figure}

All the evidence is consistent with Bootes III being a dwarf galaxy (or
a remnant thereof). Combined with its apparent location at the same
distance as Styx and very nearly in the middle of it, we infer that
Bootes III is both physically associated with the stream and quite
possibly its progenitor. Its broad spatial extent, its low surface
density and power-law profile, its possibly disturbed morphology, and
its location within Styx, all suggest that Bootes III may be in or
nearing the final throes of tidal dissolution.

\section{Constraints on Orbits} \label{orbit}

Though a lack of velocity information prevents us from tightly
constraining the orbits of the streams, the apparent orientations of
the streams combined with our distance estimates can yield some
initial constraints.  We use the Galactic model of \citet{allen91},
which assumes a spherical halo potential. We employ a least squares
method to fit both the stream orientations on the sky and our
estimated distances. The tangential velocities at each point are
primarily constrained by the projected paths of the streams while our
relative distance estimates help to limit the range of possible radial
velocities at any point.  

We fit to a number of normal points lying along the estimated
centerlines of each stream. We adopt a solar Galactocentric distance
of 8.5 kpc, and distance uncertainties of 1.0, 2.5, 2.5, and 10 kpc for
Acheron, Cocytos, Lethe, and Styx, respectively. We choose fiducial
points in each stream for radial velocity predictions based in part on
the apparent strength of the stream at those locations. When selecting
targets for follow-up spectroscopy, these regions will presumably have
the largest concentrations of stream stars and thus the highest
targeting priority.  The predicted heliocentric radial velocities and
proper motions at the fiducial points are listed in Table 1. The
uncertainties correspond to the 90\% confidence interval for each
parameter and do not take account of inaccuracies in the model
potential. Predictions are provided for both prograde
and retrograde orbits. Three-view projections of the best fit orbits
are shown in Figure 15.

\begin{figure}
\epsscale{1.0}
\plotone{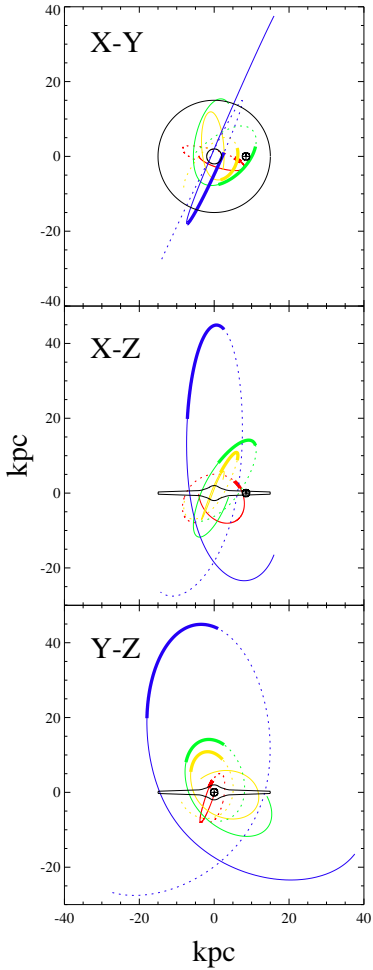}
\caption{Orbit projections for the four new streams in X, Y, Z
Galactic coordinates. The red curve corresponds to Acheron, yellow
to Cocytos, green to Lethe, and Styx is represented by the blue
curve. The heavy lines show the portions of the orbits visible in
Figure 1. The thin solid curves show the orbits 
integrated forward in time while the dotted portions of the curves
show the same orbit integrated backwards. The Sun's position at (X,Y,Z) =
(8.5,0,0) kpc is indicated.}
\end{figure}

Do the computed orbits suggest possible associations between the new
streams and known residents of the halo? Integrating orbits for a
sufficiently long time, one is almost certain to pass near the current
positions of known globular clusters. However, given our limited
knowledge of the Galactic potential and the rather rudimentary orbit
constraints determined above, such an exercise would clearly be
pointless. However, if we limit ourselves to integrating orbits no
more than once around the Galaxy, we can match the computed positions,
distances, and radial velocities against the 150 globular clusters
compiled by \citet{harris96}.  Setting rejection limits of 5\arcdeg~
in sky position, 5 kpc or a 30\% difference in distance, and 50 km
s$^{-1}$ in radial velocity, we can say that there are no known
globular clusters that match our orbits simultaneously in position,
distance, and radial velocity.  Similarly, other than Bootes
III there are no known dwarf galaxies that lie along the integrated
orbit of Styx. While the uncertainties in our orbit estimates do not
allow us to make strong statements at this point, our results are
consistent with the hypothesis that the newly detected globular
cluster streams are remnants of clusters that have long since
been whittled away to nothing.

\section{Conclusions} \label{conclusions}

Applying optimal contrast filtering techniques to SDSS data, we have
detected four new stellar streams in the Galactic halo, along with a
probable dwarf galaxy in possibly the final throes of tidal
disruption.  Three of the streams are spatially very narrow and are
likely to be the remnants of extant or disrupted globular clusters.
The fourth stream is much broader, similar in width to the Orphan and
Anticenter streams, and presumably constitutes the tidal debris of a
dwarf galaxy.  Bootes III, the new dwarf galaxy candidate, appears to
lie near the middle of this stream and may well be its
progenitor. Based on an apparently good match to the color-magnitude
distribution of stars in M 13 and M 15, we conclude that the stars
making up the streams are old and metal poor, with the Styx stream and
Bootes III being the oldest and/or most metal poor.

A determination of the nature and the properties of our dwarf galaxy
candidate with deep imaging are the subject of a forthcoming paper
\citep{grill2008}. If indeed Bootes III is in the final throes of
tidal dissolution, it will be a particularly interesting target for
detailed kinematic studies. Refinement of the stream orbits will require radial
velocity measurements of individual stars, though given the very low
stellar surface densities in these streams, this will necessarily be
an ongoing task. In this respect, these streams may be
particularly well suited for upcoming spectroscopic survey instruments such
as LAMOST.

\acknowledgments

The author is grateful to an anonymous referee for numerous
recommendations that greatly improved both the presentation and the
quality of the results. Funding for the SDSS and SDSS-II has been
provided by the Alfred P. Sloan Foundation, the Participating
Institutions, the National Science Foundation, the U.S. Department of
Energy, the National Aeronautics and Space Administration, the
Japanese Monbukagakusho, the Max Planck Society, and the Higher
Education Funding Council for England. The SDSS Web Site is
http://www.sdss.org/.

The SDSS is managed by the Astrophysical Research Consortium for the
Participating Institutions. The Participating Institutions are the
American Museum of Natural History, Astrophysical Institute Potsdam,
University of Basel, University of Cambridge, Case Western Reserve
University, University of Chicago, Drexel University, Fermilab, the
Institute for Advanced Study, the Japan Participation Group, Johns
Hopkins University, the Joint Institute for Nuclear Astrophysics, the
Kavli Institute for Particle Astrophysics and Cosmology, the Korean
Scientist Group, the Chinese Academy of Sciences (LAMOST), Los Alamos
National Laboratory, the Max-Planck-Institute for Astronomy (MPIA),
the Max-Planck-Institute for Astrophysics (MPA), New Mexico State
University, Ohio State University, University of Pittsburgh,
University of Portsmouth, Princeton University, the United States
Naval Observatory, and the University of Washington.

{\it Facilities:} \facility{Sloan}.

\begin{deluxetable}{lrrcccccccccc}
\tablecaption{Predicted Motions and Orbit Parameters}
\tablecolumns{11}
\tablewidth{0pc}
\tablehead{
\multicolumn{1}{c} {Stream} &
\multicolumn{2}{c} {Fiducial Point} &
\multicolumn{3}{c} {Prograde Orbit} &
\multicolumn{3}{c} {Retrograde Orbit} \\
\multicolumn{1}{c} {}&
\multicolumn{1}{c} {R.A.} &
\multicolumn{1}{c} {dec} &
\multicolumn{1}{c} {$v_r$} &
\multicolumn{1}{c} {$\mu_\alpha$ cos($\delta$)} &
\multicolumn{1}{c} {$\mu_\delta$} &
\multicolumn{1}{c} {$v_r$} &
\multicolumn{1}{c} {$\mu_\alpha$ cos($\delta$)} &
\multicolumn{1}{c} {$\mu_\delta$} &
\multicolumn{1}{c} {$R_{peri}$} &
\multicolumn{1}{c} {$R_{apo}$} \\
\multicolumn{1}{c} {} & 
\multicolumn{2}{c} {J2000} &
\multicolumn{1}{c} {km s$^{-1}$} &
\multicolumn{1}{c} {mas yr$^{-1}$} &
\multicolumn{1}{c} {mas yr$^{-1}$} &
\multicolumn{1}{c} {km s$^{-1}$} &
\multicolumn{1}{c} {mas yr$^{-1}$} &
\multicolumn{1}{c} {mas yr$^{-1}$} &
\multicolumn{1}{c} {kpc} &
\multicolumn{1}{c} {kpc}}
\startdata
Acheron & 15 50 24 & +9 48 39 & $-240^{+99}_{-78}$  & $-3.5 \pm 0.5$ &
$-4.2 \pm 0.3$ & $136^{+104}_{-74}$ & $-12.5 \pm 0.5$ & $-13.3 \pm 0.3$ & $3.5 \pm 0.8 $& $9.2 \pm 3.3 $\\
Cocytos  & 16 29 21 & +26 40 8 & $ -142 \pm 10$ & $0.51 \pm 0.07$ &
$-3.80 \pm 0.05$ & $-92 \pm 10$ & $-5.8 \pm 0.1$ & $-1.0 \pm 0.05$ & $ 4.9 \pm 0.2 $ & $12.5 \pm 0.2 $\\
Lethe  & 16 15 45 & +29 56 45 & $ -134 \pm 13$ & $+0.67 \pm 0.19$ & $ -3.46 \pm 0.04$ & $-105 \pm 13$ & $ -5.1 \pm 0.2$ & $ -0.40 \pm 0.03$ & $7.7 \pm 0.4$ & $17.3 \pm 0.5 $\\
Styx & 13 56 24 & +26 48 00  &  $ -42^{+44}_{-191}$ & $ -0.23^{+0.09}_{-0.17}$ & $ -0.6 \pm 0.03$ & $-22^{+182}_{-44}$ & $-1.05 \pm 0.09$ & $ -0.95 \pm 0.03$ &$13.7 \pm 3.9$ & $45^{+38}_{-1}$\\
\enddata
\end{deluxetable}

\clearpage



\end{document}